\documentclass[twocolumn,showpacs,preprintnumbers,superscriptaddress,amsmath,amssymb]{revtex4-1}

\usepackage{epsfig}
\usepackage{graphicx}% Include figure files
\usepackage{dcolumn}% Align table columns on decimal point
\usepackage{bm}% bold math
\usepackage{times}
\usepackage{xcolor}

\begin{document}

%\preprint{APS/123-QED}

\title{Predicting the epidemic threshold of the
  susceptible-infected-recovered model}

\author{Wei Wang}
\affiliation{Web Sciences Center, University of Electronic
Science and Technology of China, Chengdu 610054, China}

\affiliation{Big data research center, University of Electronic
Science and Technology of China, Chengdu 610054, China}

\affiliation{Center for Polymer Studies and Department of Physics,
  Boston University, Boston, Massachusetts 02215, USA}

\author{Quan-Hui Liu}
\affiliation{Web Sciences Center, University of Electronic
Science and Technology of China, Chengdu 610054, China}
\affiliation{Big data research center, University of Electronic
Science and Technology of China, Chengdu 610054, China}

\author{Lin-Feng Zhong}
\affiliation{Web Sciences Center, University of Electronic
Science and Technology of China, Chengdu 610054, China}
\affiliation{Big data research center, University of Electronic
Science and Technology of China, Chengdu 610054, China}

\author{Ming Tang}  \email{tangminghuang521@hotmail.com}
\affiliation{Web Sciences Center, University of Electronic
Science and Technology of China, Chengdu 610054, China}
\affiliation{Big data research center, University of Electronic
Science and Technology of China, Chengdu 610054, China}

\author{Hui Gao}
\affiliation{Web Sciences Center, University of Electronic
Science and Technology of China, Chengdu 610054, China}
\affiliation{Big data research center, University of Electronic
Science and Technology of China, Chengdu 610054, China}

\author{H. Eugene Stanley}
\affiliation{Center for Polymer Studies and Department of Physics,
  Boston University, Boston, Massachusetts 02215, USA}

\date{\today}

\begin{abstract}

Researchers have developed several theoretical methods for predicting
epidemic thresholds, including the mean-field like (MFL) method, the
quenched mean-field (QMF) method, and the dynamical message passing
(DMP) method. When these methods are applied to predict epidemic threshold
they often produce differing results and their relative levels of accuracy are
still unknown. We systematically analyze these two issues---relationships
among differing results and levels of accuracy---by studying the
susceptible-infected-recovered (SIR) model on uncorrelated  configuration
networks and a group of 56 real-world networks. In uncorrelated
 configuration networks the MFL and DMP methods yield identical predictions
that are larger and more accurate than the prediction generated by the
QMF method. When compared to the 56 real-world networks, the epidemic
threshold obtained by the DMP method is closer to the actual epidemic
threshold because it incorporates full network topology information and
some dynamical correlations. We find that in some scenarios---such as
networks with positive degree-degree correlations, with an eigenvector
localized on the high $k$-core nodes, or with a high level of
clustering---the epidemic threshold predicted by the MFL method, which
uses the degree distribution as the only input parameter, performs
better than the other two methods. We also find that the performances of
the three predictions are irregular versus modularity.

\end{abstract}

%\pacs{89.75.Hc, 87.19.X-, 87.23.Ge}
\maketitle

\small

Because many real-world phenomena incorporate spreading dynamics on
complex networks, the topic has received much attention over the last
decade~\cite{Pastor-Satorras2014, Castellano2009}. Notable examples
include the spread of sexually-transmitted diseases through contact
networks~\cite{Rocha2010}, the spread of malware on wireless
networks~\cite{Hu2009}, and the spread of computer viruses through email
networks~\cite{Newman2002b}. In each case the spreading dynamics are
strongly affected by network topology, and this complicates the task of
understanding their behavior. Existing studies of spreading dynamics
have focused on both theoretical aspects (e.g., nonequilibrium critical
phenomena~\cite{Dorogovtsev2008,Moreno2001}) and practical issues (e.g.,
proposing efficient immunization
strategies~\cite{Cohen2003,Wang2014a}). Researchers have focused on
developing ways of accurately identifying epidemic thresholds because of
their important ramifications in many real-world
scenarios. Theoretically speaking, an epidemic threshold characterizes
the critical condition above which a global epidemic
occurs~\cite{Moreno2001}. Being able to predict an epidemic threshold
allows us to determine the critical exponents~\cite{Mata2015} and
Griffiths effects~\cite{Munoz2010}, which are important in research on
nonequilibrium phenomena~\cite{Dorogovtsev2008}. Practically speaking,
quantifying an epidemic threshold allows us to determine the
effectiveness of a given immunization strategy~\cite{Cohen2003}. A
proposed immunization strategy is effective if it increases the epidemic
threshold.  In addition, knowing the epidemic threshold enables us to
more accurately determine the optimum source node~\cite{Kitsak2010}.

Researchers have put much effort into developing a theory for
quantifying the thresholds in epidemic spreading models such as the
susceptible-infected-recovered (SIR) model~\cite{Pastor-Satorras2014}.
The best-known theoretical methods fall into three categories based on
the topology information that they use. The first is the mean-field like
(MFL) approach, which uses the degree distribution as the sole input
parameter. This category includes the heterogeneous mean-field
theory~\cite{Moreno2001,Pastor-Satorras2001}, the percolation
theory~\cite{Newman2002}, the edge-based compartmental
approach~\cite{Volz2001, Miller2012,Wang2014,Wang2015}, and the pairwise
approximation method~\cite{Eames2002,Gross2006}. The second type is the
quenched mean-field (QMF) method that describes network topology in terms of the
adjacent matrix. Examples include the discrete-time Markov
chain~\cite{Gomez2010} and the $N$-intertwined
approach~\cite{Mieghem2011}.  The third type is the dynamical
message passing (DMP) method~\cite{Karrer2011} that describes network topology
in terms of the non-backtracking matrix. This approach is accurate
in the case of tree-like networks.  Researchers have used these three
approaches to uncover the macroscopic statistical characteristics (e.g.,
degree~\cite{Moreno2001} and weight distributions~\cite{Wang2014}),
mesoscale structure (e.g., degree-degree correlations~\cite{Boguna2003},
clustering~\cite{Serrano2006} and community~\cite{Newman2009}), and
microcosmic characteristics (e.g., node degree~\cite{Ferreira2012} and
edge weight~\cite{Wang2014}) that strongly affect the epidemic
threshold. For example, uncorrelated or correlated networks with a
strongly heterogeneous degree distribution can, under certain
conditions, reduce or even eliminate the epidemic
threshold~\cite{Moreno2001,Boguna2003}.

The theoretical approaches always assume (i) that an epidemic can
spread on a large, sparse network~\cite{Newman2002,Moreno2001,
Miller2012,Miller2012,Shrestha2015}, (ii) that dynamical correlations
among the neighbors do not exist~\cite{Moreno2001}, and (iii) that all
the nodes or edges within a given class are statistically
equivalent~\cite{Moreno2001, Wang2014}. These three methods also usually
focus on a class of networks, such as uncorrelated networks, clustering
networks, and community networks. In any given network the three
theoretical methods usually predict differing epidemic thresholds.  To
determine the relationships among the three differing outcomes of
the MFL method, the QMF method, and the DMP method and to
determine which more closely describes real-world epidemic thresholds,
we use a comprehensive study of the SIR model on uncorrelated
configuration networks and of a group of 56 real-world networks.
We find that the MFL and DMP methods predict
the same epidemic threshold value for uncorrelated configuration
networks and that this value is larger and more accurate than the value
predicted by the QMF method. The relationships among the three
theoretical predictions for real-world networks, however, remain
unclear.  In the 56 real-world networks studied, the DMP method performs
the best because it considers the full topology and many of the
dynamical correlations among the states of the neighbors, but due to the
localized eigenvector of the adjacent matrix the QMF method often
deviates from accurate epidemic threshold values. For networks with an
eigenvector localized on the high $k$-core nodes, positive degree-degree
correlations, or high clustering, the prediction by MFL method is more
accurate than the predictions from other two methods, even though the
MFL method uses the degree distribution as the sole input parameter. For
networks with an eigenvector localized on the hubs, negative
degree-degree correlations, or low clustering, the DMP method performs
the best. Finally, we note that the performances of the three
predictions do not exhibit an obvious regularity versus the modularity,
and in most cases the DMP method performs better than other two.

\section*{RESULTS}

\emph{\textbf{Theoretical predictions of epidemic threshold.}}
In the SIR pattern of the spread of disease though a network, at any
given time each node is either susceptible, infected, or recovered. A
susceptible node does not transmit the disease. Infected nodes
contract the disease and spread it to their neighbors. A recovered node
has returned to health and no longer spreads the disease.  To initiate
the epidemic, we randomly select a ``seed'' node and designate all other
nodes susceptible.  At each time step, infected nodes transmit the
disease to susceptible neighbors with a probability $\beta$.  Infected
nodes can also recover with a probability $\gamma$. The spreading
terminates when all infected nodes have recovered.  The spreading
dynamics can be characterized by the effective spreading rate
$\lambda=\beta/\gamma$. When $\lambda$ is below the epidemic threshold
$\lambda_c$ (i.e., $\lambda\leq\lambda_c$), the disease spreads locally
(i.e., only a tiny fraction of nodes transmit the disease).  Epidemics
can occur when $\lambda>\lambda_c$ (i.e., when a finite fraction of
nodes transmit the disease).

The mean-field like (MFL) method, the quenched mean-field (QMF) method,
and the dynamical message passing (DMP) method are commonly-used
theoretical methods of predicting an epidemic threshold.  In this
section we clarify the relationships among these epidemic thresholds
predicted by the three theoretical methods.

The mean-field like (MFL) method incorporates the heterogeneous
mean-field theory, percolation theory, the edge-based compartmental
approach, and the pairwise approximation method. Here the epidemic
threshold is predicted by using only the degree distribution, and it is
assumed that (i) all the nodes and edges in a given class are
statistically equivalent, (ii) the states of nodes among neighbors are
independent, and (iii) the network size is infinite. Using the degree
distribution $P(k)$ as the only input parameter, the theoretical
epidemic threshold prediction using the MFL method is
\begin{equation} \label{lambda_1}
\begin{split}
\lambda_c^{\rm MFL}=\frac{\langle k\rangle}{\langle
  k^2\rangle-\langle k\rangle},
\end{split}
\end{equation}
where $\langle k\rangle$ and $\langle k^2\rangle$ are the first and
second moments of the degree distribution, respectively. Although
$\lambda_c^{\rm MFL}$ is a good predictor of the epidemic threshold in
uncorrelated networks, the prediction may fail in real-world networks
because of their complex structure (e.g., degree-degree correlations,
clustering, and community) and the strong dynamical correlations among
the states of neighbors~\cite{Ferreira2012, Castellano2010}.

The quenched mean-field (QMF) method
\cite{Chakrabarti2008,Mieghem2009,Gomez2010} takes into account the
complete network structure by using the adjacent matrix $A$. This
distinguishes it from the MFL method, which simply uses the degree
distribution.  The adjacent matrix $A$ is also used to describe network
topology by the discrete-time Markov chain~\cite{Gomez2010}, the
$N$-intertwined method~\cite{Mieghem2011}, and other similar methods,
and thus they fall into the same class as the QMF method. The QMF method
is unable to capture the dynamical correlations among the states of
neighbors and uses only the correlation between the theoretical
epidemic threshold and the leading eigenvalue of the adjacent matrix to
predict the epidemic threshold, i.e.,
\begin{equation} \label{lambda_2}
\begin{split}
\lambda_c^{\rm QMF}=\frac{1}{\Lambda_A},
\end{split}
\end{equation}
where the leading eigenvalue of the adjacent matrix is~\cite{Mieghem2011}
\begin{equation} \label{leading_A}
\begin{split}
\Lambda_A={\rm max}_{\vec{v}} (\frac{\vec{v}^TA\vec{v}}
       {\vec{v}^T\vec{v}}),
\end{split}
\end{equation}
where $\vec{v}$ is a column vector with $N$ elements, and $N$ is the
network size.

The dynamical message passing (DMP) method was recently developed and
used to study nonreversible epidemic spreading dynamics in an SIR
modeled finite-sized network
\cite{Shrestha2015,Lokhov2015,Karrer2011}. The DMP method uses the
non-backtracking matrix to determine the complete network
structure. This method can both describe the complete network structure
and capture some of the dynamical correlations among the states of
neighbors that are neglected in the MFL and QMF methods. In large
sparse networks the DMP method provides a good estimation of the
epidemic threshold, i.e.,
\begin{equation} \label{lambda_3}
\begin{split}
\lambda_c^{\rm DMP}=\frac{1}{\Lambda_M},
\end{split}
\end{equation}
where
\begin{equation} \label{leading_M}
\begin{split}
\Lambda_M={\rm max}_{\vec{w}}(\frac{\vec{w}^TM\vec{w}}
       {\vec{w}^T\vec{w}})
\end{split}
\end{equation}
is the leading eigenvalue of the non-backtracking matrix
\cite{Martin2014,Krzakala2013,Karrer2014,Radicchi2015}
\begin{equation} \label{non-backtracking}
M=\left(
  \begin{array}{cc}
    A & \textbf{1}-D\\
    \textbf{1} & \textbf{0}\\
  \end{array}
\right),
\end{equation}
and $\textbf{1}$ is a $N\times N$ unit matrix, $D$ is the diagonal
matrix with the vertex degrees along its diagonal, and $\textbf{0}$ is a
$N\times N$ null matrix.

The three theoretical predictions of epidemic threshold are closely
correlated. In any given network they distinct, e.g., $\lambda_c^{\rm
  QMF}$ is less than $\langle k\rangle/\langle
k^2\rangle$~\cite{Kitsak2010}. To determine other relationships among
the three theoretical thresholds, we assume that $\lambda$ is a
eigenvalue of non-backtracking matrix $M$ and that
$w=(\overrightarrow{w_1}, \overrightarrow{w_2})^T$ is the corresponding
eigenvector of $\lambda$, where $\overrightarrow{w_1}$ and
$\overrightarrow{w_2}$ are the first and last $N$ elements of vector
$w$, respectively. Using Eq.~(\ref{non-backtracking}), the eigenvalue
problem is written
\begin{equation} \label{sep}
\left\{\begin{matrix}
A\overrightarrow{w_1}+(\textbf{1}-D)\overrightarrow{w_2}=
\lambda\overrightarrow{w_1},\\
\overrightarrow{w_1}=\lambda\overrightarrow{w_2}.
\end{matrix}\right.
\end{equation}
Multiplying the left vector $\overrightarrow{u}=(1,\cdots,1)$ on the
first line of (\ref{sep}) yields
\begin{equation}\label{lambda}
\lambda=\frac{\overrightarrow{d}^T\overrightarrow{w_1}}
{\overrightarrow{u}\overrightarrow{w_1}}-1,
\end{equation}
where $\overrightarrow{d}=(d_1,\cdots,d_N)^T$ and $d_i$ is the degree of
node $i$. In uncorrelated networks the nonbacktracking centrality of a
node is proportional to its degree~\cite{Martin2014}, i.e., $w_{1_i}\sim
d_{i}$. Here the theoretical prediction $\lambda_c^{\rm DMP}$ using the
DMP method is the same as $\lambda_c^{\rm MFL}$ using the MFL method.

To examine the eigenvalue relationships between the adjacent matrix and
non-backtracking matrix, we insert the second equation of (\ref{sep})
into the first equation and obtain
\begin{equation}\label{com}
\lambda A\overrightarrow{w_2}+(\textbf{1}-D)\overrightarrow{w_2}
=\lambda^2\overrightarrow{w_2}.
\end{equation}
Multiplying $\overrightarrow{w_2}^T$ on both sides of Eq.~(\ref{com})
and dividing $\overrightarrow{w_2}^T\overrightarrow{w_2}$, we get
\begin{equation}\label{com_con}
\frac{\lambda \overrightarrow{w_2}^T
  A\overrightarrow{w_2}}{\overrightarrow{w_2}^T\overrightarrow{w_2}}+
\frac{\overrightarrow{w_2}^T(\textbf{1}-D)\overrightarrow{w_2}}
{\overrightarrow{w_2}^T\overrightarrow{w_2}}
=\lambda^2.
\end{equation}
Using matrix theory~\cite{Mieghem2011} we know that the eigenvalue
$\epsilon$ and its corresponding eigenvector $\overrightarrow{h}$ of a
matrix $\mathcal{X}$ satisfy $\varepsilon=\frac{ \overrightarrow{h}^T
  \mathcal{X}\overrightarrow{h}}{\overrightarrow{h}^T\overrightarrow{h}}$.
We assume that $\xi_1$ and $\xi_2$ are the eigenvalue of $A$ and
$\textbf{1}-D$, respectively, i.e., $\xi_1=\frac{ \overrightarrow{w_2}^T
  A\overrightarrow{w_2}}{\overrightarrow{w_2}^T\overrightarrow{w_2}}$
and $\xi_2=\frac{ \overrightarrow{w_2}^T
(\textbf{1}-D)\overrightarrow{w_2}}{\overrightarrow{w_2}^T\overrightarrow{w_2}}$.
Thus Eq.~(\ref{com_con}) can be written as
\begin{equation} \label{rewrite}
\lambda^2=\lambda\xi_1+\xi_2.
\end{equation}
Because the minimum eigenvalue of $\textbf{1}-D$ is $1-k_{\rm max}$, we find
that
\begin{equation} \label{inequation}
\lambda^2\leq\lambda\xi_1+1-k_{\rm max}.
\end{equation}
Rewriting Eq.~(\ref{inequation}) we get
\begin{equation} \label{inequation_f}
\lambda+\frac{k_{\rm max}-1}{\lambda}\leq\xi_1.
\end{equation}
Note that $\lambda$ and $\xi_1$ are the eigenvalues of matrixes $M$ and
$A$ respectively, and we get
\begin{equation}\label{compare}
\lambda_c^{\rm DMP}\geq\lambda_c^{\rm QMF}.
\end{equation}

Many real-world networks have a heterogeneous degree distribution, e.g.,
a power-law degree distribution $P(k)\sim k^{-\nu_D}$, where $\nu_D$ is
the degree exponent.  In uncorrelated
scale-free networks, $\lambda_c^{\rm MFL}$ vanishes in the thermodynamic
limit when $\nu_D<3$ because $\langle k^2\rangle$ diverges.  When $\nu_D>3$, $\lambda_c^{\rm MFL}$ is a finite
value. Using the QMF method, the epidemic threshold $\lambda_c^{\rm
  QMF}$ is determined by the maximum degree $k_{\rm max}$. When the
degree exponent $\nu_D>2.5$, and $\lambda_c^{\rm QMF}\propto
1/\sqrt{k_{\rm max}}$.  When $\nu_D<2.5$, $\lambda_c^{\rm QMF}\propto
\langle k \rangle/\langle k^2\rangle$, which indicates that
$\lambda_c^{\rm QMF}<\lambda_c^{\rm MFL}$.  Note that $\lambda_c^{\rm
  DMP}=\langle k\rangle/(\langle k^2\rangle-\langle k\rangle)$ for
uncorrelated networks~\cite{Krzakala2013} is the same with
$\lambda_c^{\rm MFL}$. According to Eq.~(\ref{compare}), $\lambda_c^{\rm
  DMP}$ is always larger than $\lambda_c^{\rm QMF}$. Unfortunately, the
complex topology of the real-world networks makes the relationships
among the three types of prediction unclear.

\textbf{\emph{Simulation results.}}
Increasing the amount of network topology information
utilized in any predictive method, the intuitional understanding
tells us that the better performance of the
method. Using the assumptions listed in previous section, we expect
the DMP method to outperform the QMF method and the QMF method to
outperform the MFL method. We next evaluate the performance of the three
types of method using (i) a large number on SIR studies of uncorrelated
configuration networks, and (ii) 56 real-world networks.
We employ the estimators supplied in previous section to determine
the theoretical epidemic threshold, and use the relative
variance to determine the accurate epidemic threshold (see details
in Method).

To better understand the performance of the three types of method, we
further classify the networks into two classes according to the distinct
eigenvector localizations of the leading eigenvalue of the adjacent
matrix~\cite{Pastor-Satorras2015}, i.e., (i) localized hub networks
(LHNs) in which the leading eigenvalue of the adjacent matrix
$\Lambda_A$ is closer to $\sqrt{k_{\rm max}}$ than $\langle
k^2\rangle/\langle k\rangle$, where $k_{\rm max}$ is the maximum degree
of the network (the eigenvector is localized on the hub nodes), and (ii)
localized $k$-core networks (LKNs) in which $\Lambda_A$ is closer to
$\langle k^2\rangle/\langle k\rangle$ than $\sqrt{k_{\rm max}}$ (the
eigenvector is localized on nodes with a high $k$-core index).

\begin{figure}
\begin{center}
\epsfig{file=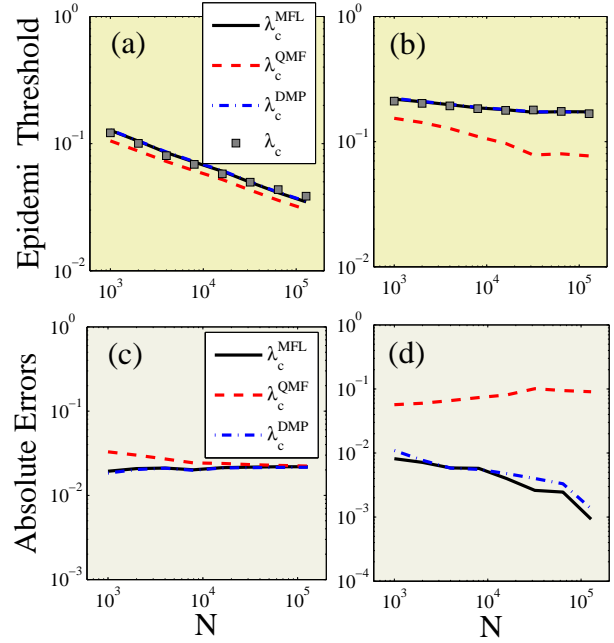,width=1\linewidth}
\caption{(Color online) {\bf Predicting epidemic threshold for
    uncorrelated configuration networks under different network sizes.}
  Theoretical predictions of $\lambda_c^{\rm MFL}$ (black solid lines),
  $\lambda_c^{\rm QMF}$ (red dashed lines), $\lambda_c^{\rm DMP}$ (blue
  dash-dotted lines) and numerical prediction (gray squares) versus
  network size $N$ for degree exponent $\nu_D=2.1$ (a) and $\nu_D=3.5$
  (b). The absolute errors between $\lambda_c$ and $\lambda_c^{\rm MFL}$
  (black solid lines), $\lambda_c^{\rm QMF}$ (red dashed lines) and
  $\lambda_c^{\rm DMP}$ (blue dash-dotted lines) versus $N$ for
  $\nu_D=2.1$ (c) and $\nu_D=3.5$ (d).}
\label{fig1}
\end{center}
\end{figure}

\emph{\textbf{Uncorrelated configuration networks.}}
Figure~\ref{fig1} shows a systematic study of the SIR model on
uncorrelated configuration networks. We focus on size $N$ scale-free
networks with power-law degree distributions, i.e., $P(k)\sim
k^{-\nu_D}$, where $\nu_D$ is the degree exponent. The minimum degree is
$k_{\rm min}=3$, and the maximum degree $k_{\rm max}$ is set at
$\sqrt{N}$, which ensures that there will be no degree-degree
correlations. Two values, $\nu_D=2.1$ and $\nu_D=3.5$, are
considered. According to definition~\cite{Pastor-Satorras2015}, networks
with $\nu_D=2.1$ are LKNs and networks with $\nu_D=3.5$ are
LHNs. Figure~\ref{fig1} shows that predictions from the MFL
($\lambda_c^{\rm MFL}$) and DMP ($\lambda_c^{\rm DMP}$) methods in
general produce similar theoretical values and perform better than the
prediction from the QMF ($\lambda_c^{\rm QMF}$) method. When
$\nu_D=2.1$, the absolute errors in the epidemic threshold from the MFL
and DMP methods are very small for all values of $N$, and the absolute
errors from the QMF method decrease with $N$. The absolute error for
method $u\in\{\mathrm{MFL},\mathrm{QMF},\mathrm{DMP}\}$ is
$\Delta(\lambda_c^{u})=|\lambda_c^u-\lambda_c|$. When $\nu_D=3.5$, the
absolute error from the QMF method stabilizes to finite values even in
infinitely large networks, and the absolute errors for the MFL and DMP
methods decrease with $N$. From these results we find that the
performance of the QMF method is counterintuitive, i.e., that its
performance is even worse than the MFL method. At the same time, all of
these results confirm the relationships among the three theoretical
predictions for uncorrelated networks previously discussed.

\textbf{\emph{Real-world networks.}}
We now examine the performances of the three theoretical predictions
$\lambda_c^{\rm MFL}$, $\lambda_c^{\rm QMF}$ and $\lambda_c^{\rm DMP}$
on a group of 56 real-world networks of various types, e.g., social
networks, citation networks, infrastructure networks, computer networks,
and metabolic networks. The Supporting Information supplies additional
statistical information about these real-world networks.  Note that
spreading processes are performed on giant connected clusters. At times,
for the sake of simplicity, we treat the directed networks as undirected
and the weighted networks as unweighted.

\begin{figure}
\begin{center}
\epsfig{file=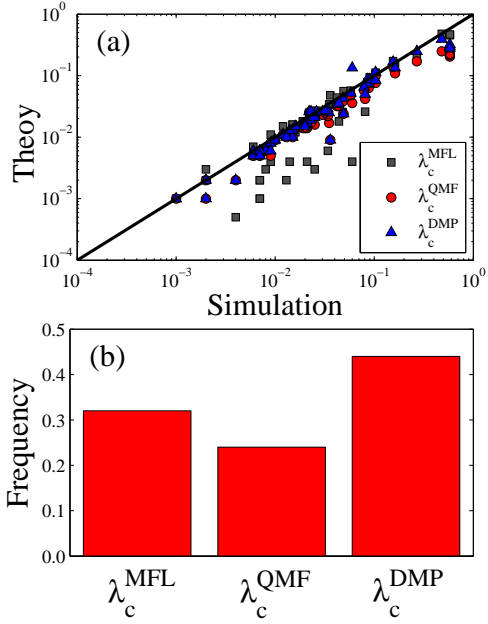,width=0.8\linewidth}
\caption{(Color online) {\bf Comparing the accuracy between three types
    of theoretical and numerical predictions of the epidemic threshold
    on $56$ real-world networks.} (a) Theoretical predictions of
  $\lambda_c^{\rm MFL}$ (gray squares), $\lambda_c^{\rm QMF}$ (red
  circles) and $\lambda_c^{\rm DMP}$ (blue up triangles) versus
  numerical predictions $\lambda_c$ of the epidemic threshold. (b) In
  all the entire sample of real-world networks, the fraction of
  $\lambda_c^{\rm MFL}$ [$\lambda_c^{\rm QMF}$ or $\lambda_c^{\rm DMP}$]
  is the closest value to $\lambda_c$.}
\label{fig2}
\end{center}
\end{figure}

Figure~\ref{fig2}(a) shows the accuracy of $\lambda_c^{\rm MFL}$,
$\lambda_c^{\rm QMF}$, and $\lambda_c^{\rm DMP}$ when applied to the 56
networks. Each symbol marks a theoretical prediction versus a numerical
network prediction.  We compute the relative frequency of $\lambda_c^{\rm
  MFL}$, $\lambda_c^{\rm QMF}$, and $\lambda_c^{\rm DMP}$ to determine
which one produces a value closest to $\lambda_c$ [see
  Fig.~\ref{fig2}(b)]. Because the DMP method considers the full
information of network topology and also some dynamical correlations,
$\lambda_c^{\rm DMP}$ is the best prediction in more than 40\% of the
networks. The $\lambda_c^{\rm QMF}$ value is the closest to the actual
epidemic threshold in 25\% of the networks, and the epidemic threshold
predicted by the MFL method, which uses the degree distribution as the
only input parameter, is closest to the real epidemic threshold in
about one-third of the real-world networks. Comparing these three predictions
we find that the DMP method outperforms the other two, i.e., when
determining the epidemic threshold in a general network, the DMP method
is more frequently accurate than the other two.

\begin{figure}
\begin{center}
\epsfig{file=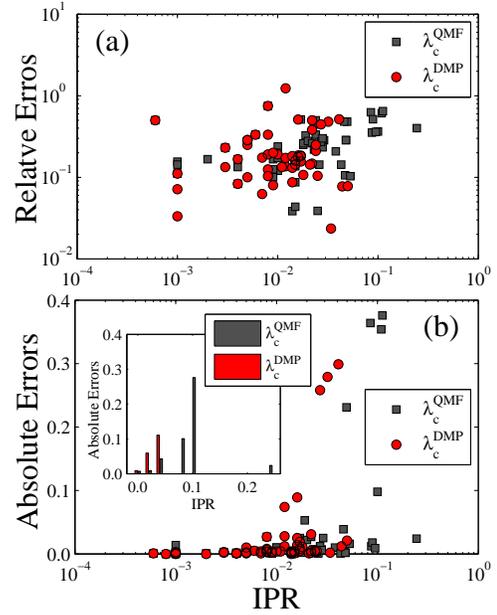,width=0.8\linewidth}
\caption{(Color online) \textbf{The effects of inverse participation
    ratio (IPR) of the adjacency and the nonbacktracking matrices on the
    accuracy of theoretical predictions.} (a) The relative errors and
  (b) absolute errors as a function of IPR of the principal eigenvectors
  of the adjacency (black squares) and the nonbacktracking matrices (red
  circles). The inset of (b) is the average absolute errors as a
  function of IPR.}
\label{fig3}
\end{center}
\end{figure}

Theoretical predictions $\lambda_c^{\rm MFL}$ given by the MFL method
often fail because it neglects much structural information and also all
dynamical correlations. The performance of the QMF method is
counterintuitive because of the localized eigenvector of the leading
eigenvalue of the adjacent matrix [see
  Fig.~\ref{fig3}(a)]. Figure~\ref{fig3} shows the effects of the
inverse participation ratios (IPR)~\cite{Goltsev2012,Karrer2014} of the
adjacent and non-backtracking matrixes. We find that the relative and
absolute errors between the theoretical and numerical predictions
increase with IPR, i.e., the QMF and DMP methods deviate from the
accurate epidemic threshold more easily when IPR is large because the
eigenvector centralities of adjacent and non-backtracking matrixes are
localized on hub nodes or high $k$-core index
nodes~\cite{Pastor-Satorras2015}. The relative error of method
$u\in\{\mathrm{MFL},\mathrm{QMF},\mathrm{DMP}\}$ can be
$\Delta^\prime(\lambda_c^u)=|\lambda_c-\lambda_c^u|/\lambda_c$.

\begin{figure}
\begin{center}
\epsfig{file=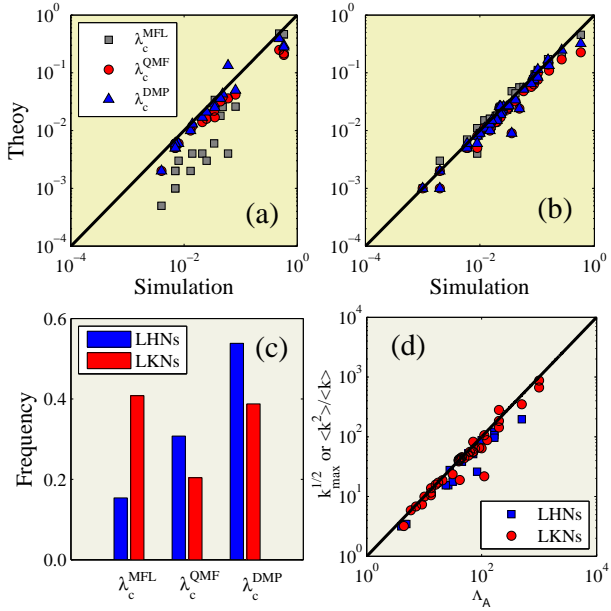,width=1\linewidth}
\caption{(Color online) \textbf{Verify the accuracy for three types of
    theoretical epidemic threshold on real-world networks.} The
  theoretical predictions of $\lambda_c^{\rm MFL}$ (gray squares),
  $\lambda_c^{\rm QMF}$ (red circles) and $\lambda_c^{\rm DMP}$ (blue up
  triangles) versus numerical predictions $\lambda_c$ of the epidemic
  threshold on (a) LHNs and (b) LKNs. (c) In the collective of LHNs and
  LKNs of real-world networks, the fraction of $\lambda_c^{\rm MFL}$
  [$\lambda_c^{\rm QMF}$ or $\lambda_c^{\rm DMP}$] is the closest value
  to $\lambda_c$. (d) The values of $k_{\rm max}^{1/2}$ for LHNs and
  $\langle k^2\rangle/\langle k\rangle$ for LKNs versus the leading
  eigenvalue $\Lambda_A$ of the adjacent matrix.}
\label{fig4}
\end{center}
\end{figure}

Recent research results indicate that networks have distinct eigenvector
localizations~\cite{Pastor-Satorras2015}. In real-world networks they
are either localized on hubs networks (LHNs) or localized on $k$-core
networks (LKNs). Depending on the localization of the eigenvector of
adjacent matrix, there are 19 LHNs and 37 LKNs among the 56 real-world
networks. Figure~\ref{fig4}(d) shows that the values $\Lambda_A$ of LHNs
are close to $k_{\rm max}^{1/2}$ (blue squares), and the values
$\Lambda_A$ of LKNs are close to $\langle k^2\rangle/\langle k\rangle$
(red circles). In LHNs [see Figs.~\ref{fig4}(a) and \ref{fig4}(c)] the
three methods perform as we would expect.  The DMP method is the best
predictor and the MFL method the worst because it neglects much detailed
network structure information. In contrast, in the LKNs [see
Figs.~\ref{fig4}(b) and \ref{fig4}(c)], the simple MFL method performs
the best, and it is slightly accurate than the DMP method.

\begin{figure}
\begin{center}
\epsfig{file=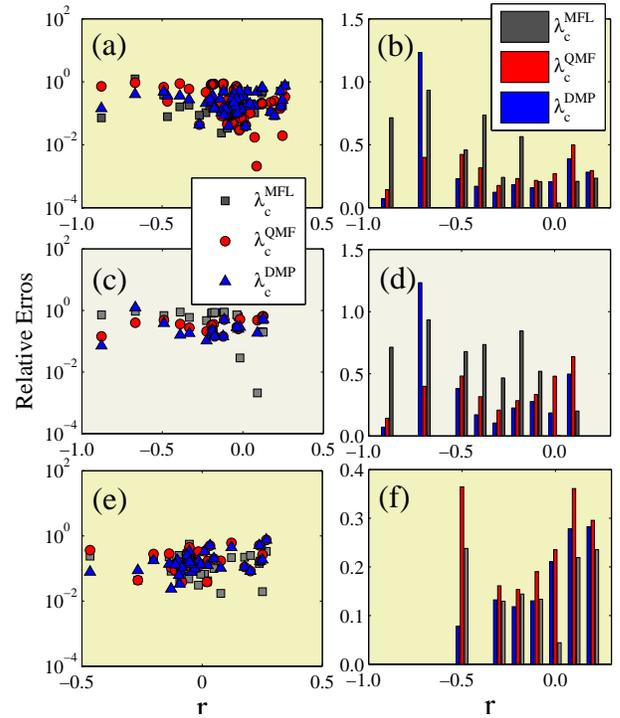,width=1\linewidth}
\caption{(Color online) \textbf{Effects of degree-degree correlations on
    the relative errors of different theoretical predictions}. In the
  first column, figures (a), (c) and (e) are the the relative errors of
  the three different theoretical predictions versus degree-degree
  correlations $r$. In the second column, figures (b), (d) and (f) are
  the the average relative errors for the three different theoretical
  predictions versus $r$. The first row exhibits the results of $56$
  real-world networks, the second row shows the results of LHNs, the
  third row performs the results of the LKNs.}
\label{fig5}
\end{center}
\end{figure}

We now compare the accuracy between the three theoretical epidemic thresholds
under different microscopic and mesoscale topologies of real-world structures,
including degree-degree correlations $r$, clustering $c$, and modularity
$Q$. To measure the accuracy of the three methods in each theoretical
prediction, we compute the average relative errors in the interval
$(x-\Delta x/2,x+\Delta x/2)$, where $x$ is $r$, $c$, and $Q$. Here we
set $\Delta x=0.1$ unless otherwise specified.  Figures~\ref{fig5}(a)
and \ref{fig5}(b) show that in all cases except the Facebook (NIPS)
network the DMP method has a lower relative error when the Pearson
correlation coefficient value is $r<0$. The Facebook (NIPS) network may
be an exception because the IPR value of its non-backtracking matrix is
relatively large, i.e., 0.012. When $r<0$, we can conclude that the
DMP method performs the best and the MFL method performs the worst.
When $r>0$, the MFL method is the most accurate and the QMF
method is the least.  Figures~\ref{fig5}(c)--\ref{fig5}(f) show the 56
real-world networks, separating them according to eigenvector
localization. In LHNs we see a phenomenon similar to that shown in
Figs.~\ref{fig5}(a) and \ref{fig5}(b), i.e., when $r<0$ the DMP method
is the most accurate and the MFL method is the least, but when $r>0$ the MFL
method is the most accurate and the QMF method is the least. In LKNs, when
$r<0$ the DMP method is the most accurate, when $r>0$ the MFL method is
the most accurate, and the QMF method is always the least accurate. This
suggests that the MFL method is the best for predicting epidemic
thresholds in networks with positive degree-degree correlations, but
that the DMP method is better in all other cases.

\begin{figure}
\begin{center}
\epsfig{file=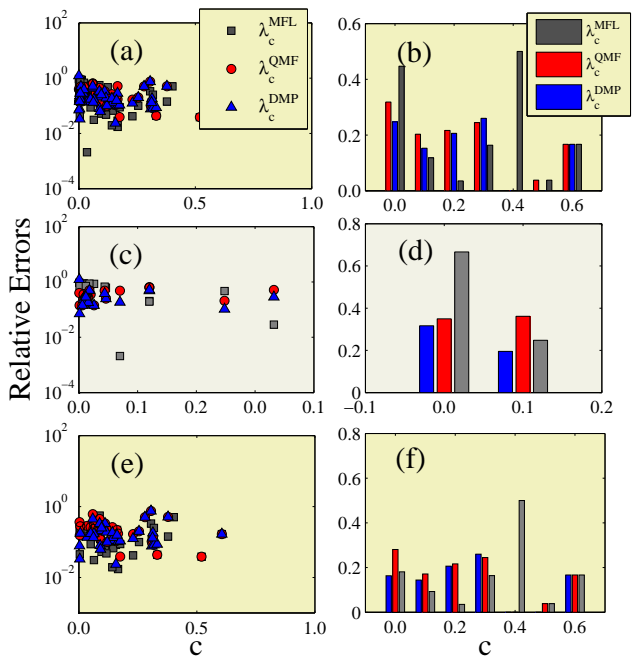,width=1\linewidth}
\caption{(Color online) \textbf{Effects of clustering on the relative
    errors of different theoretical prediction.} In the first column,
  figures (a), (c) and (e) are the the relative errors of the three
  different theoretical predictions versus clustering $c$. In the second
  column, figures (b), (d) and (f) are the the average relative errors
  for the three different theoretical predictions versus $c$. The first
  row exhibits the results of $56$ real-world networks, the second row
  shows the results of LHNs, the third row performs the results of the
  LKNs. }
\label{fig6}
\end{center}
\end{figure}

Using an analytic framework similar to that shown in Fig.~\ref{fig5},
Fig.~\ref{fig6} compares the accuracy between the three theoretical
predictions under different clustering coefficient $c$.  Figures~\ref{fig6}(a)
and \ref{fig6}(b) show that when $c<0.1$, the relative error of the DMP
method is the lowest and the relative error of the MFL method is the
largest.  When $c>0.1$, the relative error of the MFL method is the
lowest and the relative error of the QMF method is, in most cases, the
largest. Thus when $c<0.1$ the DMP method is the most accurate in predicting
the epidemic threshold, but when $c>0.1$ the MFL method is the most
accurate. In LHNs, we find the same phenomena as shown in
Figs.~\ref{fig6}(a) and \ref{fig6}(b). The DMP method is the best
predictor when $c<0.1$, and the MFL method the best when $c>0.1$ [see
  Figs.~\ref{fig6}(c) and \ref{fig6}(d)].  Figures~\ref{fig6}(e) and
\ref{fig6}(f) show that in LKNs the DMP method performs the best for
small $c$ and the MFL method the best for large $c$.

\begin{figure}
\begin{center}
\epsfig{file=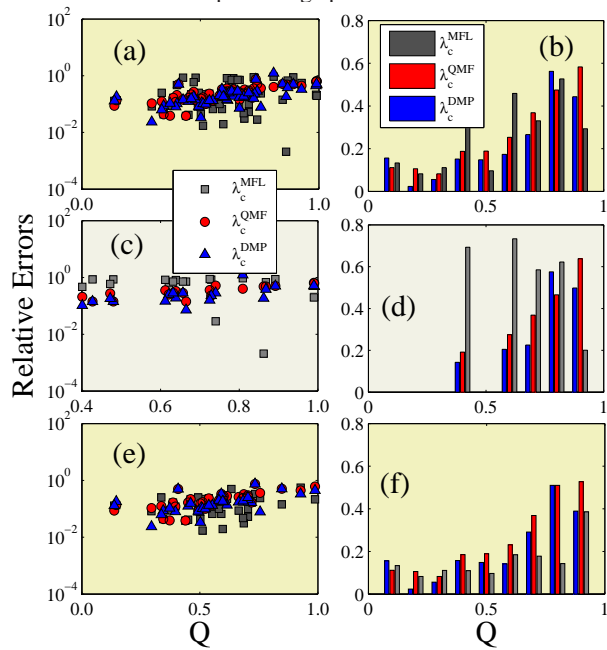,width=1\linewidth}
\caption{(Color online) \textbf{Effects of modularity on the relative
    errors of different theoretical prediction}. In the first column,
  figures (a), (c) and (e) are the the relative errors of the three
  different theoretical predictions versus modularity $Q$. In the second
  column, figures (b), (d) and (f) are the the average relative errors
  for the three different theoretical predictions versus $Q$. The first
  row exhibits the results of $56$ real-world networks, the second row
  shows the results of LHNs, the third row performs the results of the
  LKNs. }
\label{fig7}
\end{center}
\end{figure}

Finally, Fig.~\ref{fig7} compares the effectiveness between the three
predictions under different modularity $Q$. Note that in real-world networks
the relative errors increase with $Q$.  In the 56 networks, in LHNs, and
in LKNs, we note that the performances of the three predictions do
not exhibit an obvious regularity versus the modularity, and in most
cases the DMP method performs better than other two.

\section*{Conclusions}

In this study we have systematically examined the accuracies and
relationships among the MFL, QMF, and DMP methods for predicting the
epidemic threshold in the SIR model. To do this we have focused on a
large number of artificial network simulations and on 56 real-world
networks.  We first analyzed the differences and correlations among the
three theoretical epidemic threshold predictions. Generally speaking, the
three predictions differ, and the epidemic threshold predicted by the
DMP method is often larger than that predicted by the QMF method. In
uncorrelated networks, the DMP and MFL methods produce the same epidemic
threshold prediction, which is larger than the prediction produced by
the QMF method. When applied to real-world networks, however, the
relationships among the three predictions are still unclear.

We then checked the accuracies of the three predictive methods using
uncorrelated configuration networks, and found that the MFL and DMP
methods perform well, but that the QMF method does not. In the group of
56 real-world networks we found that the DMP method performs the best,
and that the epidemic threshold predicted by the MFL method is more
accurate than the one predicted by the QMF method. In networks with an
eigenvector localized on high $k$-core nodes, i.e., LKNs, the MFL method
performs the best and the QMF method the worst, but in networks with an
eigenvector localized on hubs, i.e., LHNs, the DMP method performs the
best and the MFL method the worst.

Finally we measured the performances of the three methods versus the
microscopic and mesoscale topologies in real-world networks, including
degree-degree correlations, clustering, and modularity. In networks with
negative degree-degree correlations, we found that the DMP method
performs the best, and the QMF method is better than the MFL method. In the
networks with positive degree-degree correlations, the MFL method is the
most accurate, and the QMF method is the least. In networks with low
clustering, the DMP method is the most accurate, and the MFL method is the
least.  In networks with high clustering, the MFL method is the most
accurate, and the QMF method is the least.  The relative accuracies of the
three predictions versus the modularity are, unfortunately, irregular.

Predicting accurate epidemic thresholds in networks is profoundly
significant in the field of spreading dynamics. Our results present a
counterintuitive insight into the use of network information in
theoretical methods, i.e., the performance level of a method is not only
proportional to the topological information used, but also correlates
with the dynamic correlations among the state of neighbor nodes. Our
results expand our understanding of epidemic thresholds and provide ways
of determining which method of theoretical prediction is best in a
variety of given situations.  Our results also indicate directions for
further research into the development of more accurate theoretical
methods of predicting epidemic thresholds.

\section*{METHODS}
\emph{\textbf{Predicting numerical threshold}}.
To determine the theoretical epidemic threshold, we employ the
estimators supplied by the MFL, QMF and DMP methods and use the relative
variance $\chi$ to numerically determine the size-dependent epidemic
threshold~\cite{Chen2014},
\begin{equation}\label{ch}
\chi=\frac{\langle r -\langle r\rangle\rangle^2}{r^2},
\end{equation}
where $r$ denotes the final epidemic size and $\langle\cdots\rangle$ is
the ensemble averaging.  We use at least $10^5$ independent dynamic
realizations on a network to calculate the average value of $\chi$,
which exhibits a maximum value at the epidemic threshold
$\lambda_c$. This numerical prediction $\lambda_c$ obtained by observing
$\chi$ we consider the accurate epidemic threshold~\cite{Chen2014}.  The
Supporting Information supplies illustrations of numerically locating
the epidemic threshold by observing $\chi$. There are also other ways of
determining $\lambda_c$, e.g., susceptibility~\cite{Ferreira2012} and
variability methods~\cite{Shu2015}.

\acknowledgments

This work was partially supported by the National Natural Science
Foundation of China under Grants Nos.~11105025, 11575041 and 61433014
the Program of Outstanding Ph. D. Candidate in Academic Research by
UESTC under Grand No.~YXBSZC20131065.

\end{document}